\documentclass[nofootinbib,twocolumn,showpacs,preprintnumbers,amsmath,amssymb]{revtex4}
\usepackage{graphicx}
\usepackage{dcolumn}
\usepackage{bm}
\usepackage{wasysym}

\newcommand{\mnras}{MNRAS}

\newcommand{\aap}{A\&A}

\newcommand{\jcap}{JCAP}

\DeclareMathOperator{\arcosh}{arcosh}

\begin{document}

\preprint{APS/123-QED}

\title{The Hubble stream near a massive object: the exact analytical solution for the spherically-symmetric case}

\author{A. N. Baushev}
\email{baushev@gmail.com}

\affiliation{Bogoliubov Laboratory of Theoretical Physics, Joint Institute for Nuclear Research\\
141980 Dubna, Moscow Region, Russia}

\date{\today}

\begin{abstract}
The gravitational field of a massive object (for instance, of a galaxy group or cluster) disturbs
the Hubble stream, decreasing its speed. Dependence $v(r_0)$ of the radial velocity of the stream
from the present-day radius $r_0$ can be directly observed and may provide valuable information
about the cluster properties. We offer an exact analytical relationship $v(r_0)$ for a
spherically-symmetric system.
\end{abstract}


\maketitle

\section{Introduction}
In an ideal Friedmann's universe, all nearby objects move off radially from an observer with the
speed proportional to the distance to the object (the famous Hubble-Lema\^\i tre law). The real
Universe, however, contains local overdensities (like galaxy groups or clusters), and their
additional gravitational attraction slows down the Hubble flow, changing the dependence $v(r_0)$ of
the radial velocity from the present-day radius $r_0$. Hereafter we will consider a single
overdensity and name it the 'galaxy cluster', though our reasoning is equally valid for any
cosmological object, on condition that the suppositions that we will make during the derivation are
applicable for the object. The dependence $v(r_0)$ can be obtained from observations (see, for
instance, Figure~1 in \cite{makarov2009})) and, in principle, may tell a lot about the cluster and
its environment. We just need to build a theoretical model of $v(r_0)$ as a function of cluster
parameters and compare it with observations.

Of course, the task is well-known, and there is a vast literature dedicated to its solution. The
methods can be roughly divided into three groups. The analytical approach was applied in old papers
(for example, \cite{olson1979, lb1981, giraud1986}), but the cosmological constant was believed to
be zero at that time. Several approximative formulas were offered for $v(r_0)$ (for instance,
\cite{peirani2008, makarov2009}). Recently N-body simulations are performed to solve the problem
(e.g., \cite{hanski2001, penarrubia2014}), which allowed to consider realistic non-spherical models
of the Local Group.

On the other hand, the N-body simulations may not be considered as having no disadvantages
\cite{17, 20, 21}, and the analytical approach seems preferable in simple cases, since it allows to
obtain a precise and compact equation for $v(r_0)$. For instance, \cite{24} offered an exact
analytical equation for the radius $L_0$, at which a spherically-symmetric object of mass $M$ stops
the Hubble flow\footnote{In \cite{24} $L_0$ is denoted by $R_0$.}, i.e., $v(L_0)=0$. We use the
same approach in this letter, and our aim is to generalize the result of \cite{24} and find the
full velocity profile $v(r_0)$ in the spherically-symmetric case. As we will see, one may find an
exact analytical solution of this task, even for an arbitrary (but spherically-symmetric)
distribution of matter around the cluster.

Let us specify the Universe model, which is implied in this Letter. We suppose that the Universe is
homogeneous and isotropic, i.e., its metrics can be represented as $ds^2=c^2dt^2-a^2(t) dl^2$,
where $dl$ is an element of three-dimensional length and $a$ is the scale factor of the Universe.
We denote the present-day values of the Hubble constant, critical density, and the scale factor of
the Universe by $H_0$, $\rho_{c,0}$, and $a_0$, respectively. We denote the present-day matter,
dark energy, and curvature\footnote{The curvature density is equal to $\rho_{a,0} =\dfrac{3
c^2}{8\pi G} \cdot \dfrac{k}{a_0^2}$, where $k=-1, 1, 0$ if the universe density is higher,
smaller, or equal to the critical one, respectively.} densities of the Universe by $\rho_{M,0}$,
$\rho_{\Lambda,0}$, $\rho_{a,0}$, respectively. We may also introduce the ratios of the present-day
densities of the universe components to $\rho_{c,0}$: $\Omega_{a,0}\equiv \rho_{a,0}/\rho_{c,0}$
etc. In the literature, the quantities $\Omega_{\Lambda,0}$, $\Omega_{M,0}$ etc. are often denoted
simply by $\Omega_{\Lambda}$, $\Omega_{M}$ etc., but we use the additional sub-index to remind that
these are the present-day values.

We perform our calculations for the case of the standard $\Lambda$CDM Universe (though they can be
easily generalized for less standard cosmological models). We suppose that the dark energy behaves
simply as the cosmological constant (i.e., $p_\Lambda=-\rho_{\Lambda,0}=\it{const}$), and that the
Universe is flat ($\Omega_{a,0}=0$) in absence of structures \cite{pdg18}. We neglect the
contribution of the radiation component: the present-day density of radiation is $\sim
10^{-4}\rho_{c,0}$, and, though the contribution was much larger in the early Universe, we will
show that the relative error of the velocity field determination caused by the disregarding of
radiation is also $\sim 10^{-4}$. Then we obtain: $\Omega_{\Lambda,0}+\Omega_{M,0}=1$. The universe
age $t_0$ in the $\Lambda$CDM is defined by the well-known equation (see e.g. \cite[eqn.
4.29]{gorbrub1}):
\begin{eqnarray}
t_0 H_0 \sqrt{\Omega_{\Lambda,0}} = \frac{2}{3}\arcosh\left(1/\sqrt{\Omega_{M,0}}\right)
\label{25a4}
\end{eqnarray}
We may obtain this equation from (\ref{25a5}), if we substitute there
$\sigma_{M,0}=\Omega_{M,0}\rho_{c,0}$, $\sigma_{\Lambda,0}=\Omega_{\Lambda,0}\rho_{c,0}$,
$\sigma_{a,0}=0$ and integrate.

\section{The idea of the solution}

We find the velocity field around a cluster of galaxies under the following assumptions:
\begin{enumerate}
     \item \label{as1} The system is spherically-symmetric and has not experienced any tidal
     perturbations from other structures.
     \item \label{as2} Its characteristic radius (for instance, we may consider $R_0$) is much
     larger than its gravitational radius and much smaller than the universe radius ($c/H_0\gg R_0\gg
     R_g\equiv{2GM}/{c^2}$). The significance of this assumption will be explained below.
\end{enumerate}
We choose the center of symmetry of the system as the origin of coordinates, and the Big Bang as
the zero point of time. We denote the present-day radii and radial speeds of the objects around the
cluster by $r_0$ and $v$. We emphasize that $v$ is the speed of the Hubble stream, it is refined
from the peculiar velocities of the objects. Our aim is to find the relationship $v(r_0)$.

Consider a spherical layer of radius $r_0$ and determine its previous evolution $r(t,r_0)$. Our
solution is based on two well-known properties\footnote{See \cite[chapters 1, 4]{zn2} for a
detailed proof. A brief outline of it may be found in \cite{24}.} of spherically-symmetric
gravitating systems in the general theory of relativity: first, a spherically-symmetric
distribution outside a radius $r$ does not create any gravitational field inside this radius. It
means that the matter outside $r$ does not affect the dependence $r(t,r_0)$ at all.

Second, $r(t,r_0)$ depends only on the total mass of dark and baryonic matter inside $r$, and does
not depend on the matter space distribution, if the distribution is spherically-symmetric. Indeed,
the exterior layers do not create any gravitational field at $r$, and, in accordance with the
Birkhoff's theorem, the gravitational field created at $r$ by the spherical, nonrotating matter
inside $r$ must be Schwarzschild, i.e., it depends on the only parameter, the total energy inside
$r$. Contrary to the newtonian gravity, the gravitational field in the general theory of relativity
is created by both, density and pressure. However, the only component with significant pressure in
our system is the dark energy, which has exactly the same pressure and density
everywhere\footnote{It needs not be true for more exotic models of the dark energy. For instance,
if the dark energy has some additional interaction with baryonic matter, our consideration is not
valid.}. The matter pressure is negligible with respect to its density. Therefore, the total mass
inside $r$ does not depend on the matter distribution inside $r$, and we may redistribute the
matter inside $r$ whatever we like without any influence on the gravitational force at $r$ and the
layer evolution $r(t,r_0)$.

\section{Calculations}

Thus, we may virtually redistribute the matter inside $r$ uniformly, and it will not affect the
dependence $r(t,r_0)$. But then we have a 'uniform universe' inside $r$, and its evolution may be
found from the usual Friedmann equation (see e.g. \cite[eqn. 4.1]{gorbrub1}):
\begin{equation}
\left(\dfrac{dr}{r dt}\right)^2 =\dfrac{H^2_0}{\rho_{c,0}}
\left[\sigma_{M,0}\left(\dfrac{r_0}{r}\right)^3\!\!+\sigma_{\Lambda,0}+\sigma_{a,0}\left(\dfrac{r_0}{r}\right)^2\right].
\label{25a5}
\end{equation}
Here $\sigma_{M,0}$, $\sigma_{\Lambda,0}$, $\sigma_{a,0}$ are the averaged present-day matter, dark
energy, and curvature densities inside $r_0$. Of course, $\sigma_{M,0}$, $\sigma_{\Lambda,0}$,
$\sigma_{a,0}$ depend on $r_0$. By analogy with $\Omega_{0}$, we may introduce the fractions
$\Sigma_{M,0}\equiv \sigma_{M,0}/\rho_{c,0}$ etc. We purposefully use $\sigma$ and $\Sigma$ in
order to distinguish the averaged densities $\sigma$ and $\Sigma$, depending on $r_0$, from $\rho$
and $\Omega$, that correspond to the undisturbed Universe and therefore are universal. Obviously,
$\sigma_{\Lambda,0}=\rho_{\Lambda,0}$, $\Sigma_{\Lambda,0}=\Omega_{\Lambda,0}$. Denoting the matter
mass as a function of $r_0$ by $M_M(r_0)$, we obtain
\begin{equation}
\Sigma_{M,0}(r_0)=\dfrac{3 M_M(r_0)}{4\pi r^3_0 \rho_{c,0}}. \label{25a6}
\end{equation}
Here we used \emph{assumption}~\ref{as2} about the cluster of galaxies: it implies that the space
near the cluster is flat. At $r=r_0$ $dr/dt = v(r_0)$, and we obtain from (\ref{25a5})
\begin{eqnarray}
\nonumber \left(\dfrac{v}{r_0 H_0}\right)^2 = \left[\Sigma_{M,0}+\Omega_{\Lambda,0}+\Sigma_{a,0}\right]\\
\Sigma_{a,0}= \left(\dfrac{v}{r_0 H_0}\right)^2 - \Sigma_{M,0}-\Omega_{\Lambda,0}\label{25a7}
\end{eqnarray}
We substitute this value to (\ref{25a5}) and obtain
\begin{eqnarray}
\label{25a8}\left(\frac{dx}{H_0 dt}\right)^2 =\frac{\Sigma_{M,0}}{x}(1-x) -
\Omega_{\Lambda,0}(1-x^2)+\left(\frac{v}{r_0 H_0}\right)^2,
\end{eqnarray}
where $x\equiv r/r_0$. We introduce two new designations
\begin{equation}
\alpha=\frac{\Sigma_{M,0}}{\Omega_{\Lambda,0}};\qquad U=\frac{v^2_0}{r^2_0
H^2_0\Omega_{\Lambda,0}}. \label{25a9}
\end{equation}
As we can see, $U\ge 0$. Strictly speaking, $\alpha$ and $U$ are functions of $r_0$, but now it is
more convenient for us now to consider them as independent variables. We may rewrite (\ref{25a8})
as
\begin{equation}
H_0 \sqrt{\Omega_{\Lambda,0}}dt =\frac{\sqrt{x}dx}{\sqrt{x^3+(U-\alpha-1)x+\alpha}}\equiv
\xi(U,\alpha,x) dx. \label{25a10}
\end{equation}
Here we introduced a new function $\xi(U,\alpha,x)$
\begin{equation}
\xi(U,\alpha,x)\equiv\dfrac{\sqrt{x}}{\sqrt{x^3+(U-\alpha-1)x+\alpha}} \label{25a10v2}
\end{equation}
for brevity sake. Variables $\alpha$ and $U$ do not depend on $x$, and we may easily integrate
equation~(\ref{25a10}). If $v\ge 0$ (obviously, it means that $r_0\ge L_0$), the integration is
trivial:
\begin{equation}
t_0 H_0 \sqrt{\Omega_{\Lambda,0}} =\int_0^1 \xi(U,\alpha,x) dx. \label{25a11}
\end{equation}
Comparing this equation with (\ref{25a4}), we equate their right parts:
\begin{equation}
\frac{2}{3}\arcosh\left(1/\sqrt{\Omega_{M,0}}\right) =\int_0^1 \xi(U,\alpha,x) dx, \quad \text{if }
v\ge 0. \label{25a12}
\end{equation}
This equation defines $U$ as an implicit function of $\alpha$ and $\Omega_{M,0}$. The function can
be resolved with respect to $U$ in form of an explicit function $U_+(\alpha, \Omega_{M,0})$. The
subscript '$+$' reminds that the function describes only the positive velocity branch, or $r_0\ge
L_0$. We should underline that equation~(\ref{25a12}) gives a meaningful relation $U_+(\alpha,
\Omega_{M,0})$ not for all pairs of $U$ and $\alpha$. This issue will be discussed in
section~\ref{25discussion}.

\begin{figure}
    \resizebox{\hsize}{!}{\includegraphics{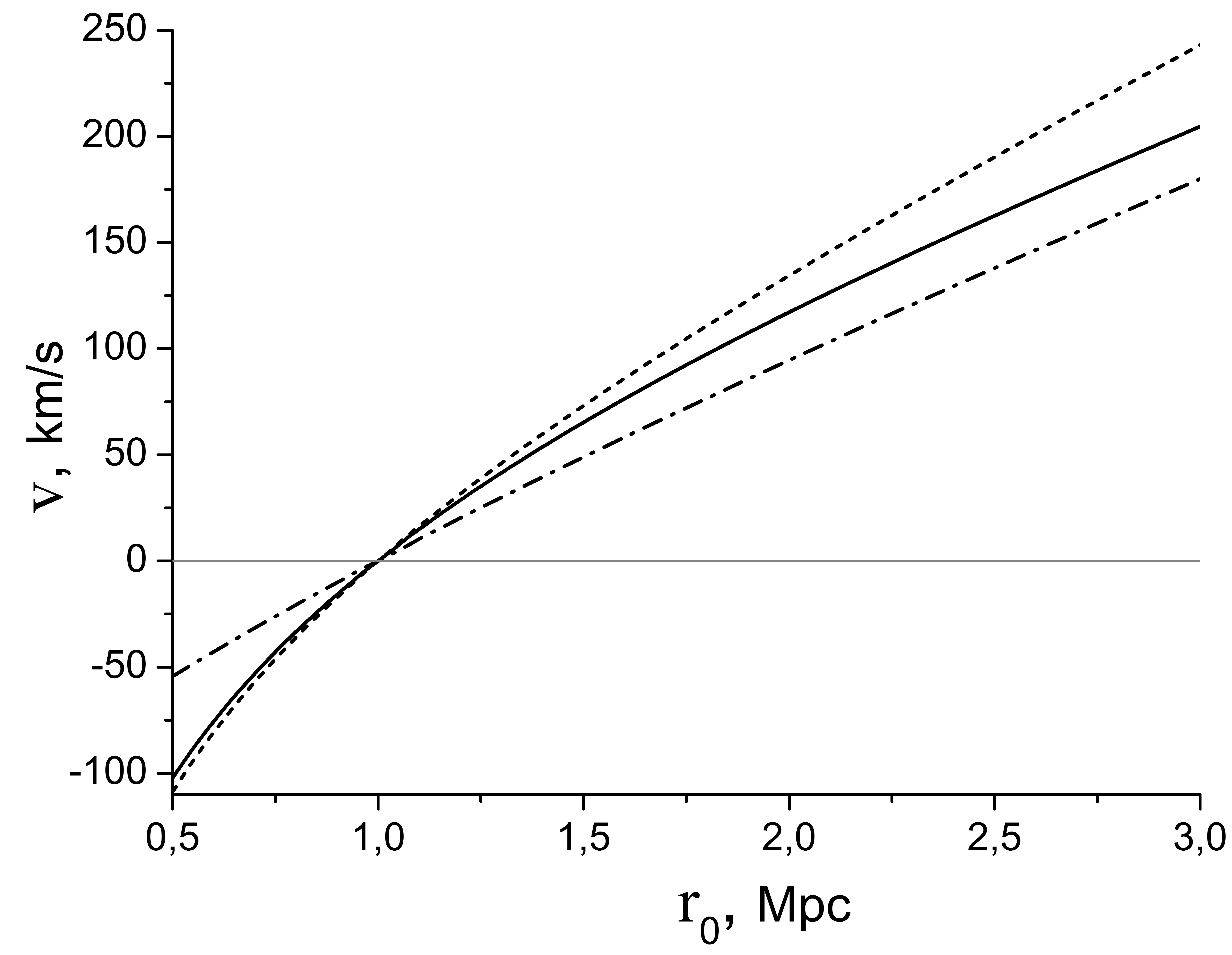}}
    \vspace{-0.3cm}
    \caption{The speed profiles of the Hubble stream around the cluster for the \emph{point} model~(\ref{25a20}) (solid line),
    \emph{halo} model~(\ref{25a21}) (dash-dot line), and \emph{empty} model~(\ref{25a22}) (dot line), respectively, at the distances
    $0.5-3$~{Mpc} from the cluster center.} \label{25fig1}
    \vspace{-0.3cm}
\end{figure}

If we substitute $U=0$ into (\ref{25a12}), we obtain an implicit function, bounding the average
matter density inside the stop radius $L_0$, namely $\alpha_0 =\Sigma_{M,0}/\Omega_{\Lambda,0}$,
with $\Omega_{M,0}$:
\begin{equation}
\frac{2}{3}\arcosh\left(1/\sqrt{\Omega_{M,0}}\right) =\int_0^1
\frac{\sqrt{x}dx}{\sqrt{x^3-(\alpha_0+1)x+\alpha_0}}. \label{25a13}
\end{equation}
This equation coincides with equation (11) from \cite{24} and allows to find $L_0$ for an arbitrary
spherically-symmetric mass distribution $M(r_0)$ (see \cite{24} for details).

The integration of (\ref{25a10}) is more complex, if $v < 0$ (i.e., $r_0< L_0$). In this case, the
toy 'uniform universe' inside $r$, which we consider, has already passed its maximum expansion and
now contracts (therefore $v < 0$). The maximum expansion $x_{max}=r_{max}/r_0$ corresponds to the
moment when $dr/dt=0$ (or $dx/dt=0$), i.e., the radicand in (\ref{25a10}) turns to zero. Thus
$x_{max}$ is a root of the equation $x^3+(U-\alpha-1)x+\alpha=0$.  Since $x$ varies from $0$ to
$x_{max}$, we should choose the smallest real positive one. Actually, the root should exceed $1$
(since the 'universe' contracts, its maximum radius $r_{max}$ should exceed the present-day radius
$r_0$). A cubic equation has three roots, and two of them can be complex. However, expression
$f(x)=x^3+(U-\alpha-1)x+\alpha$ is equal to $\alpha>0$ at $x=0$, and $f(x)\to -\infty$ if $x\to
-\infty$. Consequently, one of the real roots is obligatory negative and cannot be $x_{max}$. Thus,
$x_{max}$ exists only if all three roots of the equation are real. It is easy to show that we
should choose the middle one:
\begin{eqnarray}
\nonumber &x_{max}=\dfrac{\left(1-i \sqrt{3}\right) (U-\alpha-1)}{3\cdot
   2^{2/3}\cdot \left(-\alpha+\sqrt{27 \alpha^2+4 (U-\alpha-1)^3}\right)^{1/3}}-\\
   &-\dfrac{\left(1+i \sqrt{3}\right)
   \left(-\alpha+\sqrt{27 \alpha^2+4 (U-\alpha-1)^3}\right)^{1/3}}{2^{4/3}}. \label{25a15}
\end{eqnarray}
If $x_{max}$ obtained with this equation for some pair of $U$ and $\alpha$ is not real or
$x_{max}<1$, it means that $v$ cannot be negative for these values of $U$ and $\alpha$. We will
specify the necessary conditions in section~\ref{25discussion}.

Thus, if $v < 0$, the right part of (\ref{25a10}) should be integrated from $x=0$ to $x=x_{max}$,
and then from $x=x_{max}$ to $x=1$:
\begin{eqnarray}
\nonumber t_0 H_0& \sqrt{\Omega_{\Lambda,0}} =\int\limits_0^{x_{max}} \xi(U,\alpha,x) dx +
\int\limits_1^{x_{max}}\xi(U,\alpha,x) dx=\\
=&2\int\limits_0^{x_{max}} \xi(U,\alpha,x) dx - \int\limits_0^{1}\xi(U,\alpha,x) dx,  \text{if } v<
0.\label{25a16}
\end{eqnarray}
Comparing this equation with (\ref{25a4}), we equate their right parts and obtain the final
solution for the $v< 0$ case:
\begin{eqnarray}
\nonumber \frac{2}{3}\arcosh\left(1/\sqrt{\Omega_{M,0}}\right) =\\
=2\int\limits_0^{x_{max}} \xi(U,\alpha,x) dx - \int\limits_0^{1}\xi(U,\alpha,x) dx, \text{if } v<
0. \label{25a17}
\end{eqnarray}
We derive equations~(\ref{25a12}) and~(\ref{25a17}) equating the universe ages, and justifies the
disregard of the radiation term in the Friedmann equation. Radiation dominated in the early
Universe, but for less than $10^{-4}$ of the Universe age ($<5\cdot 10^5$ years). For almost all of
the $13.6$~bln. years the Universe has been living with the fraction of radiation, comparable with
the present-day one ($\Omega_{\gamma,0}< 10^{-4}$), and we may conclude that the relative error
occurring from the neglect of the radiation term does not exceed $10^{-4}$.

\begin{figure}
    \resizebox{\hsize}{!}{\includegraphics{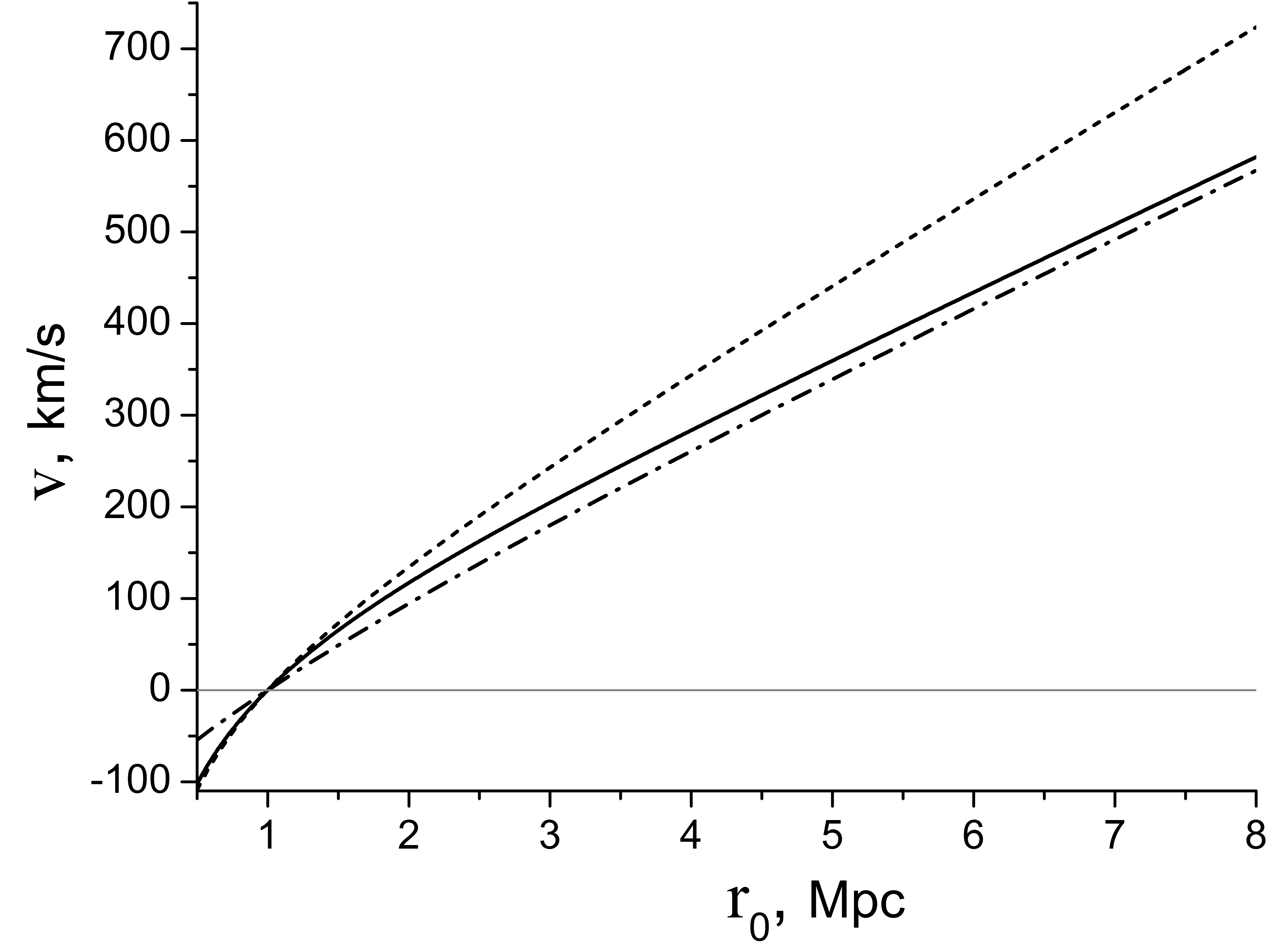}}
    \vspace{-0.3cm}
    \caption{The speed profiles of the Hubble stream around the cluster for the \emph{point} model~(\ref{25a20}) (solid line),
    \emph{halo} model~(\ref{25a21}) (dash-dot line), and \emph{empty} model~(\ref{25a22}) (dot line), respectively, at the distances
    $0.5-8$~{Mpc} from the cluster center.} \label{25fig2}
    \vspace{-0.3cm}
\end{figure}

\section{Discussion}
\label{25discussion}

Equations~(\ref{25a12}) and~(\ref{25a17}), together with definitions~(\ref{25a9})
and~(\ref{25a10v2}), give the full analytical solution of the task under consideration (the case of
$\Omega_{\Lambda,0}= 0$, is considered in the Appendix section). In order to find the speed
distribution $v(r_0)$ around an arbitrary spherically-symmetric mass distribution $M(r_0)$, we
should use the following algorithm:
\begin{enumerate}
     \item \label{alg1} We find the function $\Sigma_{M,0}(r_0)$ (equation~(\ref{25a6}))
     and $\alpha(r_0)=\Sigma_{M,0}(r_0)/\Omega_{M,0}$.
     \item \label{alg2} We find $L_0$ with the help of~(\ref{25a13}) or \cite{24}.
     \item \label{alg3} With the help of~(\ref{25a12}), we find the distribution
     $U_+(\alpha(r_0), \Omega_{M,0})$ for $r_0\ge L_0$, and with the help of~(\ref{25a17}) ---
     the distribution $U_-(\alpha(r_0), \Omega_{M,0})$ for $r_0 < L_0$.
     \item \label{alg4} With the help of~(\ref{25a9}), we restore the desired function $v(r_0)$ from
$U_+(\alpha(r_0), \Omega_{M,0})$ and $U_-(\alpha(r_0), \Omega_{M,0})$.
\end{enumerate}
It is important to underline that our solution cannot be valid for very small radii: as we derive
it, we assume that the layers with different $r_0$ do not cross each other. It is true outside of
$L_0$ and to some extend inside $L_0$. However, the stream accreting on the galaxy group finally
faces the substance that has already passed through the group and moves outwards. Inside this
radius (obviously, it lies inside $L_0$ and well outside the virial radius $R_{vir}$) we have a
multi-stream regime, and our solution fails.

Equations~(\ref{25a12}) and~(\ref{25a17}) are not defined for an arbitrary pair of $U$ and
$\alpha$. We have already discussed some limitations, but now we need to specify them in more
details.

Let us start from the $v< 0$ case, i.e., from equation~(\ref{25a17}). The convergence of the
integrals in it depends on the behavior of $f(x)=x^3+(U-\alpha-1)x+\alpha$. We have already seen
that equation $f(x)=0$ obligatory has a negative real root, and a meaningful solution of
equation~(\ref{25a17}) exists only if all three roots of the equation $x^3+(U-\alpha-1)x+\alpha=0$
are real, and the smallest positive one $x_{max}\ge 1$. Let us consider the limits, which these
conditions set on $U$ and $\alpha$. All three roots of the cubic equation are real if the
discriminant $Q\equiv [(U-\alpha-1)/3]^3+(a/2)^2\le 0$. However, the case $Q=0$ does not fit: the
cubic parabola is tangent to the $x$ axis at $x_{max}$ in this case, and we will show in the next
paragraph that then the first integral in~(\ref{25a17}) diverges at $x_{max}$. It follows from
$Q<0$ that $U<\alpha+1-3(\alpha/2)^{2/3}$. But $U\ge 0$, and thus $\alpha>2$. The physical reason
of this limitation is trivial: the gravitational attraction created by normal matter is two times
weaker than the effective gravitational repulsion created by the cosmological constant of equal
density \cite{tolman1934}. Therefore, if $\alpha\le 2$ (i.e., $\Omega_{\Lambda,0}\ge 2
\Sigma_{M,0}$), the overdensity is too low to stop the Hubble stream, and its speed $v$ cannot be
negative. Thus, we obtain the necessary conditions of meaningfulness of solution~(\ref{25a17}) for
$v< 0$:
\begin{equation}
U<\alpha+1-3(\alpha/2)^{2/3}, \qquad \alpha>2. \label{25a18}
\end{equation}
However, if we substitute the first of them into~(\ref{25a15}), we obtain after simple
transformations $x_{max}>\sqrt{\alpha/2}$. Thus, $x_{max}>1$ if $\alpha>2$, and
conditions~(\ref{25a18}) are also sufficient.

Now consider the case when $v\ge 0$, i.e., equation~(\ref{25a12}). First of all, the integral
in~(\ref{25a12}) diverges if $f(x)$ is tangent to the $X$-axis at any point $x_t\in [0,1]$. Indeed,
in this case $f(x)\propto (x-x_t)^2$ near $x_t$, the denominator in~(\ref{25a12})
$\sqrt{f(x)}\propto |x-x_t|$, and the integral diverges logarithmically. In particular, it diverges
if $\alpha=2$, $U=0$ (then $x_t=1$). The second condition is that $f(x)\ge 0$ at $x\in [0,1]$. The
conditions can be summarized as $f(x)> 0$ at $x\in [0,1]$, which is equal to
\begin{eqnarray}
\nonumber U>\alpha+1-3(\alpha/2)^{2/3}, \qquad \alpha\le 2\\
\text{any}\quad U\ge 0, \qquad \alpha> 2 \label{25a19}
\end{eqnarray}
If this condition is satisfied, the integral in equation~(\ref{25a12}) is always defined and real.

We should emphasize that we cannot guarantee that any pair ($\alpha, U$) satisfying conditions
(\ref{25a18}) or (\ref{25a19}) corresponds to a physically meaningful solution. For instance, if
$(U-\alpha-1)>0$, the layer expands with acceleration even at $x=0$, i.e., at $z\gg 10$. Of course,
it could not happen in the real Universe. However, if conditions (\ref{25a18}) and (\ref{25a19})
are not satisfied for some ($\alpha, U$), the integrals in~(\ref{25a12}) and~(\ref{25a17}),
respectively, cannot be calculated, and this pair ($\alpha, U$) is physically impossible for our
task.

To conclude this section, we should say that the task that we consider may be significantly
simplified, if we use the 'natural coordinates' $y\equiv r_0/L_0$ and $v/v_H\equiv v/(r_0 H_0)$,
where $v_H\equiv r_0 H_0$ is the speed of the undisturbed Hubble stream. Definitions~(\ref{25a9}),
together with equations~(\ref{25a12}) and~(\ref{25a17}) state that the ratio $v/v_H$ as a function
of $y$ is uniquely defined by the function $\Sigma_{M,0}\left(y\right)$, on condition that
$\Omega_{\Lambda,0}$ is the same (which is quite natural, since $\Omega_{\Lambda,0}$ is a
fundamental Universe parameter). In other words, two spherical galaxy clusters with different
$L_0$, but with the same matter distributions $\Sigma_{M,0}\left(r_0/L_0\right)$ as functions of
$r_0/L_0$, have the same distributions $\frac{v}{v_H}\left(y\right)$. Thus, the Hubble speed
distribution for a cluster with a different radius but similar matter distribution may be found by
a simple rescaling.

\begin{figure}
    \resizebox{\hsize}{!}{\includegraphics{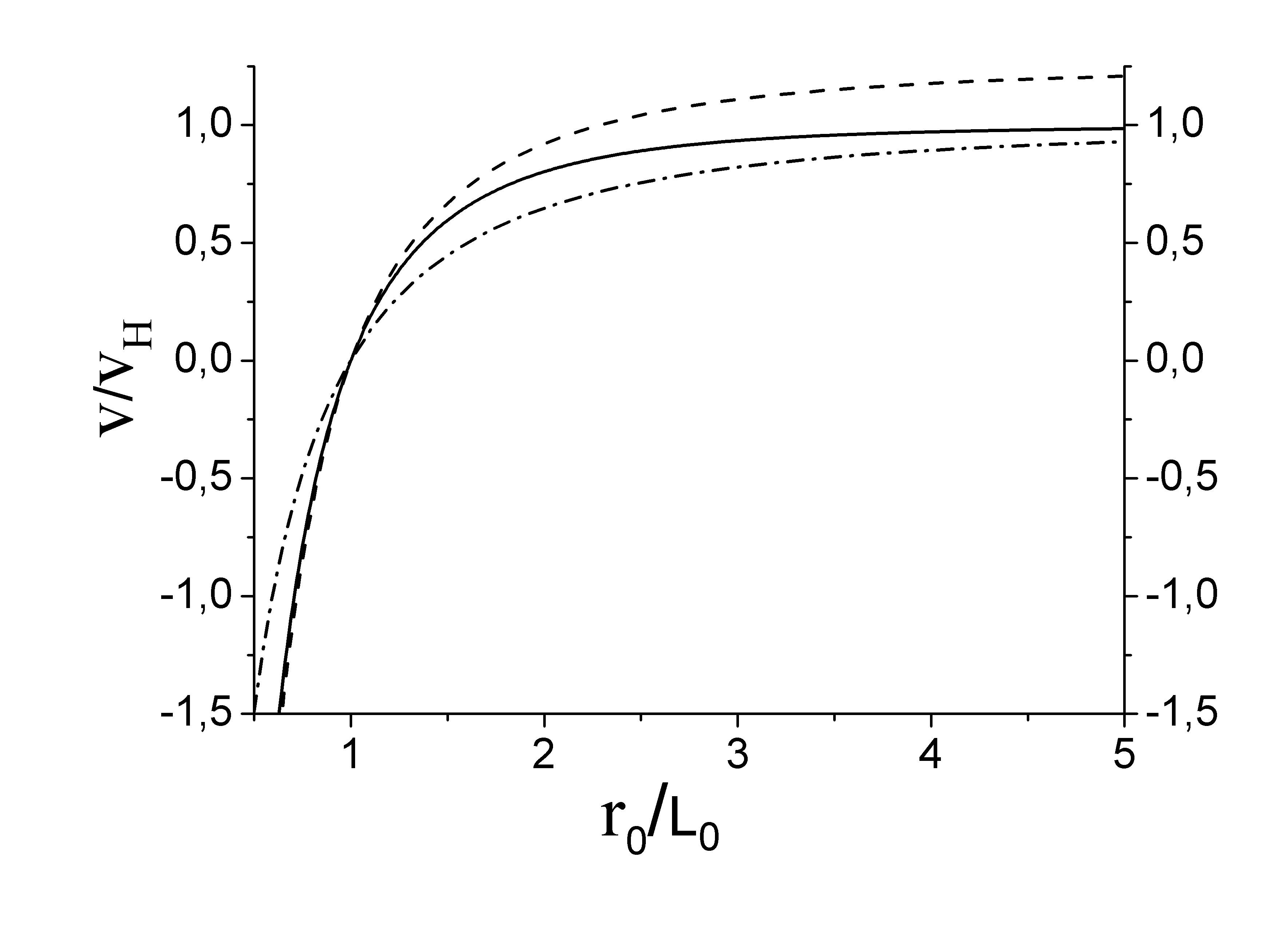}}
    \vspace{-0.3cm}
    \caption{The speed profiles of the Hubble stream around the cluster for the \emph{point} model~(\ref{25a20}) (solid line),
    \emph{halo} model~(\ref{25a21}) (dash-dot line), and \emph{empty} model~(\ref{25a22}) (dot line), respectively,
    represented in the natural coordinates $r_0/L_0$ and $v/v_H\equiv v/(r_0 H_0)$ (see the last paragraph in
    section~\ref{25discussion}).} \label{25fig3}
    \vspace{-0.3cm}
\end{figure}

\section{An example of application}
In order to illustrate the solution that we have obtained, let us calculate the velocity field for
several toy models of the galaxy clusters. We accept the Hubble constant value $H_0=73$~{km/(s
Mpc)}. In order to compare the velocity profiles for different parameters, we set all the models so
that they have the same stop radius $L_0=1$~{Mpc}, which roughly corresponds to the Local Group
value $L_0\simeq 0.9$~{Mpc} \cite{makarov2009}. As we could see in the previous paragraph, the
velocity profile for a different $L_0$ may be found by a simple rescaling. As a good illustration
of this self-similarity of the task, \cite{24} found that the average matter overdensity inside
$L_0$, $\Sigma_{M,0}(L_0)$ depends only on $\Omega_{M,0}$, and does not depend on the cluster size.
In particular, $\Sigma_{M,0}(L_0)= (3\pi/4)^2$ for $\Omega_{M,0}=1$ and $\Sigma_{M,0}(L_0)\simeq
3.67$ for $\Omega_{M,0}=0.306$, which is the value measured for our Universe \cite{pdg18}. Since
$L_0$ is the same for all the models, the matter mass inside $L_0$ is also equal. By mass and
density we mean only the matter mass and density everywhere in this section. We suppose that the
dark energy always has the uniform density distribution $\Omega_{\Lambda,0} \rho_{c,0}$.

We consider three toy models of the density profile of the cluster. First, the \emph{point} model,
where the cluster has a central point mass $M_c$ and surrounded by relatively small constant
density $\rho_M=\Omega_{M,0} \rho_{c,0}$ (which is equal to the average matter density in the
Universe). The \emph{halo} model also has two components: the uniform distribution of matter with
the density $\Omega_{M,0} \rho_{c,0}$ and a large halo with $\rho\propto r^{-2}$, the factor being
chosen so that $L_0=1$~{Mpc}. The \emph{empty} model supposes that all the matter is concentrated
in the cluster center, and the space around is empty (i.e., contains only dark energy). Contrary to
the first two models, the \emph{empty} one has incorrect asymptotic behavior: it does not transform
into the undisturb Universe at large distances, and its average density tends to
$\Omega_{\Lambda,0} \rho_{c,0}$, and not to $\rho_{c,0}$, as $r_0\to \infty$.

The matter distributions $\Sigma_{M,0}(r_0)$ (calculated from $M_M(r_0)$ with the help of
equation~(\ref{25a6})) for all three models are the following:
\begin{eqnarray}
\!\Sigma_{M,0}(r_0)\!&=&\!(\Sigma_{M,0}(L_0)-\Omega_{M,0})\left(L_0/r_0\right)^3+\Omega_{M,0}\label{25a20}\\
\!\Sigma_{M,0}(r_0)\!&=&\!(\Sigma_{M,0}(L_0)-\Omega_{M,0})\left(L_0/r_0\right)^2+\Omega_{M,0}\label{25a21}\\
\!\Sigma_{M,0}(r_0)\!&=&\!\Sigma_{M,0}(L_0)\left(L_0/r_0\right)^3\label{25a22}
\end{eqnarray}
The velocity distributions are calculated with the help of equations~(\ref{25a12})
and~(\ref{25a17}) and presented in Figures~\ref{25fig1} and~\ref{25fig2} for the \emph{point}
model~(\ref{25a20}) (solid line), \emph{halo} model~(\ref{25a21}) (dash-dot line), and \emph{empty}
model~(\ref{25a22}) (dot line), respectively. One may see that the difference between the curves is
rather significant at $r_0\sim 3$~{Mpc}, though the difference between the \emph{point} and
\emph{empty} models is not that drastic: the density $\rho_M=\Omega_{M,0} \rho_{c,0}$ of the flat
component of the former model is more than ten times lower than the average matter density
$\sigma_{M,0}(L_0)\simeq 3.67 \rho_{c,0}$ inside $L_0=1$~{Mpc}. It suggests that the matter density
profile near $L_0$ may in principle be restored from the Hubble stream observations of a galaxy
cluster.

Finally, Figure~\ref{25fig3} represents the velocity profiles for the three models in the natural
coordinates $r_0/L_0$ and $v/v_H\equiv v/(r_0 H_0)$ (see the last paragraph in
section~\ref{25discussion} for details). One can see in Figures~\ref{25fig2} and~\ref{25fig3} that
the velocity profiles corresponding to the \emph{point} and \emph{halo} models converge to the
undisturbed Hubble stream $v\to v_H=r_0 H_0$ at large radii, while the profile of the \emph{empty}
model goes significantly higher, and the ratio $v/v_H$  tends to a limit significantly larger than
$1$ in this case. It is not surprising: as we have already mentioned, asymptotically both
\emph{point} and \emph{halo} models transform into the undisturb Universe. On the other hand, the
\emph{empty} model lacks the matter, and the uncompensated effective repulsion induced by the dark
energy accelerates the Hubble expansion. The influence of the central object becomes negligible at
large distances, and the \emph{empty} model transforms into an empty Friedmann's 'universe' filled
only with the dark energy, which is apparently characterized by a new Hubble constant higher than
$H_0$. Of course, the \emph{empty} model cannot be valid at large distances in reality.

To conclude, let us note that all the preceding consideration may be easily generalized to the case
when the redshift $z$ of the galaxy cluster is not zero. It is enough just to find the matter
fraction $\Omega_{M,z}$ and the Hubble constant $H_z$ at the moment $z$, and use these values
instead of $\Omega_{M,0}$ and $H_0$, since the choice of the 'present moment' was arbitrary in our
calculations. Since $(z+1)=r_0/r$, we obtain from (\ref{25a5})
\begin{eqnarray}
\nonumber H^2_z = H^2_0 \left(\Omega_{M,0}(z+1)^3 +\Omega_{\Lambda,0}\right)\\
\Omega_{M,z} = \dfrac{\Omega_{M,0}(z+1)^3}{\Omega_{M,0}(z+1)^3+\Omega_{\Lambda,0}}.\label{25a23}
\end{eqnarray}
For instance, if $z\gg 1$, we may neglect the dark energy (i.e., accept $\Omega_{M,0}= 1$). Then
the velocity profile may be found from the equations derived in the Appendix.

We would like to thank the Heisenberg-Landau Program, BLTP JINR, for the financial support of this
work. This research is supported by the Munich Institute for Astro- and Particle Physics (MIAPP) of
the DFG cluster of excellence "Origin and Structure of the Universe".

\appendix

\section{The pure matter case ($\Omega_{\Lambda,0}=0$)}
Deriving equation~(\ref{25a10}) from (\ref{25a8}), we assume that $\Omega_{\Lambda,0}\ne 0$ and
divide by it. As a result, $\alpha$ and $U$ tend to infinity as $\Omega_{\Lambda,0}\to 0$. Thus,
the important instance of $\Omega_{\Lambda,0}= 0$ (i.e., $\Omega_{M,0}= 1$) should be considered
separately. Of course, this case does not correspond to the modern Universe, but if the cluster has
$z\gg 1$, we may neglect the dark energy. If $\Omega_{\Lambda,0}= 0$, equation~(\ref{25a8}) can be
rewritten as
\begin{equation}
\left(\dfrac{dx}{dt}\right)^2 =H^2_0
\left[\Sigma_{M,0}\left(\frac{1}{x}-1\right)+\left(\frac{v}{r_0 H_0}\right)^2\right]. \label{25x1}
\end{equation}
We may introduce a new quantity $w(r_0)=1-\dfrac{v^2_0}{r^2_0 H^2_0 \Sigma_{M,0}}$.
Since we consider an overdensity, $\Sigma_{M,0}\ge 1$, $v\le r_0 H_0$, and therefore $w\in [0;1]$.
We may rewrite the equation of motion~(\ref{25x1}) as\vspace{-0.1cm}
\begin{equation}
H_0 \sqrt{\Sigma_{M,0}} dt = \dfrac{\sqrt{x} dx}{\sqrt{1-wx}}. \label{25x3}
\end{equation}
As in the general case, the maximum expansion $x_{max}=r_{max}/r_0$ corresponds to the moment when
the radicand in~(\ref{25x3}) turns to zero, i.e., $x_{max}=1/w$. Now we may exactly follow the
derivation of equations~(\ref{25a12}) and~(\ref{25a17}), integrating~(\ref{25x3}). If $v\ge 0$, we
integrate~(\ref{25x3}) from $x=0$ to $x=1$:\vspace{-0.1cm}
\begin{equation}
H_0 t_0 \sqrt{\Sigma_{M,0}} = \dfrac{\arcsin\left(\sqrt{w}\right)-\sqrt{w(1-w)}}{w^{3/2}}.
\label{25x4}
\end{equation}
If $\Omega_{M,0}= 1$, the Universe age $t_0$ is bound with the Hubble constant by simple relation
$H_0 t_0=2/3$ \cite{zn2}. We obtain  \vspace{-0.1cm}
\begin{equation}
\frac23 \sqrt{\Sigma_{M,0}} = \dfrac{\arcsin\left(\sqrt{w}\right)-\sqrt{w(1-w)}}{w^{3/2}}, \quad
\text{if } v\ge 0. \label{25x5}
\end{equation}
If $v< 0$, we integrate~(\ref{25x3}) from $x=0$ to $x_{max}=1/w$, and then from $x_{max}$ to $x=1$.
After some trivial calculations, we obtain\vspace{-0.1cm}
\begin{equation}
\frac23 \sqrt{\Sigma_{M,0}} = \dfrac{\pi-\arcsin\left(\sqrt{w}\right)+\sqrt{w(1-w)}}{w^{3/2}},
\quad \text{if } v<0. \label{25x6}
\end{equation}
Equations~(\ref{25x5}) and~(\ref{25x6}) define $v$ as an implicit function of $r_0$ and
$\Sigma_{M,0}(r_0)$.

\end{document}